\documentclass[11pt,a4paper]{article}
\usepackage[]{amsmath,amssymb}
\usepackage{graphics,epsfig}
\begin{document}
\date{}
\title{Entanglement entropy for a Dirac fermion in three dimensions: vertex contribution}
\author{H. Casini\footnote{e-mail: casini@cab.cnea.gov.ar}, M. Huerta
\footnote{e-mail: marina.huerta@cab.cnea.gov.ar}
 and L. Leitao \\
{\sl Centro At\'omico Bariloche and Instituto Balseiro,}\\{\sl 
R8402AGP-S.C. de Bariloche, R\'{\i}o Negro, Argentina}}
\maketitle

\begin{abstract}
In three dimensions there is a logarithmically divergent contribution to the entanglement entropy which is due to the vertices located at the boundary of the region considered. In this work we find the corresponding universal coefficient for a free Dirac field, and extend a previous work in which the scalar case was treated. The problem is equivalent to find the conformal anomaly in three dimensional space where multiplicative boundary conditions for the field are imposed on a plane angular sector. As an intermediate step of the calculation we compute the trace of the Green function of a massive Dirac field in a two dimensional sphere with boundary conditions imposed on a segment of a great circle.
\end{abstract}

\section{Introduction}

A typical manifestation of entanglement is the presence of correlations for the results of measurements on commuting sets of observables. In quantum field theory, the vacuum fluctuations provide a source of quantum entanglement between spatially separated regions. A natural measure of this entanglement is the geometric entropy, which is defined as the von Neumann entropy $S(V)=-\textrm{tr}(\rho_V\log \rho_V)$ of the reduced density matrix of the vacuum state  $\rho_V=\textrm{tr}_{-V}\left|0\right\rangle\left\langle 0\right|$ corresponding to $V$. The trace in the expression for $\rho_V$ is taken over the Hilbert space generated by the degrees of freedom lying outside $V$. 

Recently, there has been an increasing interest in understanding the manifestations of entanglement in quantum field theory (QFT). This is in part due to the conjecture that black hole entropy may become geometric entropy once the role of gravity is elucidated \cite{rolegrav}. The investigation has also revealed that the entanglement entropy is a very interesting tool in the study of certain aspects of QFT. Very different phenomena such as quantum phase transitions \cite{phase}, confinement \cite{confi}, topological phases \cite{topo}, and the renormalization group irreversibility \cite{irrev}, leave an imprint on the behavior of the geometric entropy.      
 
Our hope is that this quantity, which has very powerful non perturbative properties \cite{noper}, may help to classify the different QFT, or even provide an alternative setting where QFT may be expressed in a geometrical way.  In this sense, it is remarkable that there is a concrete proposal in the context of the AdS-CFT duality where the entanglement entropy of the boundary  CFT may be computed by purely geometric means \cite{taka}. This hypothesis has been supported by further arguments \cite{argu} and has passed already several nontrivial tests \cite{tests}. 
    
With this general motivation, in this paper we study the entanglement entropy associated to a spatial polygonal set in $(2+1)$ dimensions for a free Dirac field.  The structure of divergences the entropy develops in the continuum limit is given in three dimensions by the expansion 
\begin{equation}
 S(V)=g_{1}[\partial V] \,\epsilon^{-1} + g_0[\partial V]\,\log (\epsilon \Lambda)+ S_0(V)\,,   \label{div}
\end{equation}
where $S_0(V)$ is a finite term, $\epsilon$ is a short distance cutoff, and $\Lambda$ is a parameter with dimensions of mass. 
The coefficient of the logarithmic term $g_0[\partial V]$ is a universal quantity depending on the set boundary $\partial V$~\cite{ff,anguloboson}. Because of its ultraviolet origin, for general polygonal sets we expect it to consist in a sum of contributions from the individual vertices 
\begin{equation}
g_0[\partial V]=\sum_i s(x_i)\,,
\end{equation}
where the $x_i$ denote the vertex angles. The function $s(x)$ satisfy  $s(x)=s(2 \pi-x)$, which is a consequence of the symmetry in the entropies $S(V)=S(-V)$ due to the purity of the vacuum state.

Here we find the function $s(x)\equiv s_D(x)$ for a Dirac field. In a previous work we have found some quantities related to the function $s_S(x)$ corresponding to a free scalar field \cite{anguloboson}.
 The logarithmically divergent term in the entropy due to the vertices has also been found in the holographic proposal by Ryu and Takayanagi \cite{angholo}, presumably corresponding to an interacting CFT. We will also compare our results with this last work.       
 
The method usually employed to find these entropies is based on the following representation
\begin{equation}
S(V)=\lim_{n\rightarrow 1} \frac{\log(\textrm{tr}\rho_V^n)}{(1-n)} \label{limit}\,,
\end{equation}
in terms of the trace of powers of the density matrix. Then $\rho_V^n$ is expressed as 
\begin{equation}
\textrm{tr}\rho^{n}= \frac{Z_n}{(Z_1)^{n}}\,, 
\label{dd1}
\end{equation}
where $Z_n$ is a functional integral 
 on an $n$-sheeted $d$ dimensional Euclidean space with a conical singularity of angle $2\pi n$ located at the boundary of the set $V$ (see for example \cite{see}). Here $d$ is the space-time dimension, $d=3$ in the present work. 
 The replicated space is obtained considering $n$ copies of the $d$-space cut along $V$, and sewing together the upper side of the cut in the $k^{\textrm{th}}$ copy with
the lower one of the $(k+1)^{\textrm{th}}$ copy, for $ k=1,...,n$, and where the copy $n+1$ coincides with the first one. 

In general, obtaining these integrals explicitly is a very difficult problem since we have to deal with a non trivial manifold resulting from the replication method. Moreover, the entanglement entropy follows from $Z[n]$ for integer values of $n$ through an analytic continuation to $n=1$. 
In the case of free fields, a great simplification follows by mapping the $n-$sheeted problem to an equivalent one in which we deal with $n$ decoupled free fields $\tilde{\phi}_{k}$ for $ k=1,...,n$ \cite{fermion, boson}. These fields are multivalued and live on the Euclidean $d$ dimensional space with boundary conditions imposed on the $d-1$ dimensional set $V$ given by
\begin{equation}
\tilde{\phi}^{(+)}_k(\vec{r})=e^{i\frac{2 \pi k}{n}}\tilde{\phi}^{(-)}_k(\vec{r})\,,\,\,\,\,\,\,\,\,\,\,\, \, \vec{r}\in V\,.
\label{bc}
\end{equation}
Here $\tilde{\phi}^{(+)}_k$ and $\tilde{\phi}^{(-)}_k$ are the limits of the field as the variable approaches $V$ from each of its two opposite sides in $d$ dimensions. 
In this formulation we have for fermions  
\begin{equation}
\log(\textrm{tr}\rho_V^n)=\sum_{k=-(n-1)/2}^{(n-1)/2}\textrm{log}\, Z\left[e^{i 2 \pi \frac{k}{n}}\right]\,,
\label{r2}
\end{equation}
and for bosons
\begin{equation}
\log(\textrm{tr}\rho_V^n)=\sum_{k=0}^{n-1}\textrm{log}\, Z\left[e^{i 2 \pi \frac{k}{n}}\right]\,,
\label{r1}
\end{equation}
where $Z[\lambda]$ is the partition function corresponding to a field on a single copy of the Euclidean space, which is multiplied by a factor $\lambda$ when the variable crosses $V$.  
Although $Z\left[e^{i 2 \pi k/n}\right]$ has been found explicitly  for several cases of interest \cite{anguloboson,fermion,boson} , the evaluation of the entropy in this scenario is still limited by the difficulties in getting the analytic continuation of the sum for non-integer $n$ in order to take the limit (\ref{limit}).
Only very recently \cite{analitic}, this problem has been solved by exploiting the relation between (\ref{r2}) and (\ref{r1}) with the expressions for $\rho _{V}$  in
terms of correlators for free boson and fermion systems \cite{chung,araki}.  The result is 
\begin{equation}
S=\int_0^{\infty}dt \frac{\pi}{\sinh^2(\pi t)}\log Z\left[e^{2\pi t}\right]\,,\label{vein}
\end{equation}
for free fermions, and 
\begin{equation}
S=-\int_0^{\infty}dt \frac{\pi}{\cosh^2(\pi t)}\log Z\left[-e^{2\pi t}\right]\,,\label{boti}
\end{equation}
for free bosons. Note that the functional integral involved has a boundary condition on the cut $V$ given by a real factor (instead of a phase factor as in (\ref{r2}) and (\ref{r1})), being a positive factor for fermions and a negative one for  bosons.
The calculation of the geometric entropy for the set $V$ is then effectively reduced to the one of the corresponding functional integral $Z[\lambda]$. In particular, the term logarithmic in the cutoff appears in this partition function, with a coefficient proportional to the conformal anomaly induced by the boundary conditions (\ref{bc}) \cite{vas}.

In order to obtain $s_D(x)$ we calculate the Dirac partition function for a set $V$ formed by a plane angular sector. 
We first reduce the problem to a two dimensional one on a sphere with a cut along a segment of a great circle, and then relate the Green function on this sphere to the corresponding scalar problem studied in \cite{anguloboson}. The reported results in this last work concern only the exact values for the trace of powers of the local density matrix for a scalar, and only the small angle limit for $s_S(x)$ was obtained. Here we use the formula (\ref{vein}) which allows us to give the analytic result for the entropy in terms of a system of coupled nonlinear ordinary differential equations. 

The outline of the paper is as follows. In the next Section we show how the three dimensional problem in Euclidean space can be reduced to a two dimensional one on a sphere. In the third Section, we show the correspondence between the Dirac and scalar Green functions on the cut sphere.  Then, in Section IV, we use these results and the ones of ref. \cite{anguloboson} to find the logarithmic contribution to the entropy. We conclude comparing the coefficients $s(x)$ for the scalar and fermionic cases with the data given by Hirata and Takayanagi in \cite{angholo}, corresponding to a conformal theory holographic to gravity in four dimensions.

\section{Vertex contribution to the entropy in three dimensions}
In order to extract the contribution of the vertices to the entropy, the simplest set $V$ to consider is an infinite two dimensional plane angular sector of angle $x$. We take its vertex on the origin of coordinates. 
 The associated partition function for a Dirac field  $\Psi$ in three  dimensions is  
\begin{equation}
Z[e^{i 2 \pi a}]=\int {\cal D}\Psi^\dagger {\cal D}\Psi e^{-\int dr^3 \,\Psi^\dagger  {\cal D}_3 \Psi}\,,\label{eee}
\end{equation}
where ${\cal D}_3$ is the Dirac operator in three dimensions given by
\begin{equation}
{\cal D}_3=(\gamma^{i}\partial_i+\mu)\,,
\end{equation}
 and $\gamma^{i}=\sigma^{i}$ are the Pauli matrices. The boundary condition for the spinors is 
 \begin{equation}
\Psi^+(\vec{r})=e^{i 2\pi a} \Psi^-(\vec{r})\,,\hspace{2cm} \vec{r}\in V\,.\label{bounj}
\end{equation}
Here $\Psi^+$ and $\Psi^-$ are the limit values of the field on each of the sides the two dimensional angular sector $V$ has in three dimensions. Since by reflexion symmetry on the plane containing $V$ we have $Z[e^{i 2\pi a}]=Z[e^{-i 2\pi a}]$ we can choose without loss of generality 
\begin{equation}
a\in (0, 1/2)\,.
\end{equation} 
Note that here we are using a real $a$, which allow for a direct computation of $\textrm{tr}\, \rho^n$.  In order to use the formula (\ref{vein}) to obtain the entropy we will turn to imaginary $a$ at the end of the calculation.

Due to the boundary condition, the field is singular as we approach the edges of the plane angular sector $V$ or to the vertex point. However, in order to have finite action, we have to impose that it diverges at most as
\begin{equation}
\Psi(\vec{r})\sim d^{\kappa_1} , \hspace{2cm} \kappa_1 > -\frac{1}{2}\,,\label{uneta}
\end{equation} 
as $\vec{r}$ approaches $\partial V$ (the edges of $V$), where $d$ is the distance between $\vec{r}$ and $\partial V$. Also we have  
\begin{equation}
\Psi(\vec{r})\sim |\vec{r}|^{\kappa_2}, \hspace{2cm} \kappa_2 > -1\,, \label{doseta}
\end{equation}
as we approach the origin of coordinates. The same conclusion can be reached considering that the induced current circulating around $\partial V$ has to be finite.  

The functional $Z$ is calculated exploiting the relation between the free energy and the Green function $G_D^{(3)}(\vec{r},\vec{r}^\prime)$, 
\begin{equation}
\frac{d\log Z}{d\mu}=-\int dr^3 \left< \Psi^\dagger (\vec{r}) \Psi(\vec{r})\right>=\textrm{Tr} \, G_D^{(3)}\,,\label{pero}
\end{equation}
where $\mu$ is the field mass. The Euclidean Green function $G_D$ satisfies the equation
\begin{equation}
 {\cal D}_3 G_D^{(3)}(\vec{r},\vec{r}^\prime) =\delta^3(\vec{r}-\vec{r}^\prime)\,.
\end{equation}

\subsection{Dimensional reduction}
The Dirac operator and the boundary conditions allow the separation of the angular and radial equations in polar coordinates. We will use this fact in order to reduce the problem to one in two dimensions.
In this coordinates ${\cal D}_3$ writes
\begin{equation}
	{\cal D}_3=\frac{D}{r}+\tilde{\gamma}^r\partial_r+\mu\,,
\end{equation}
with
\begin{equation} D=(\tilde{\gamma}^{\theta}\partial_{\theta}+\tilde{\gamma}^{\phi}\partial_{\phi})\,,
\end{equation}
and where the new gamma matrices $\tilde{\gamma}$ are 
\begin{equation}
\tilde{\gamma}^{\theta}=r\, \frac{\partial\theta}{\partial x_i} \sigma^i \,,\hspace{1cm}
\tilde{\gamma}^{\phi}=r\, \frac{\partial\phi}{\partial x_i} \sigma^i.\,,\hspace{1cm}
\tilde{\gamma}^{r}=\frac{\partial r}{\partial x_i} \sigma^i=-i\sin\theta\tilde{\gamma}^{\theta}\tilde{\gamma}^{\phi}\,.
\end{equation}
The operator $D$ satisfies
\begin{equation}
D\tilde{\gamma}^r+\tilde{\gamma}^{r}D=2\,, \hspace{2cm} D^2-\tilde{\gamma}^{r}D=\Delta_\Omega\,,\label{h2}
\end{equation}
where 
\begin{equation}
\Delta_{\Omega}=\frac{1}{\sin \theta}\frac{\partial}{\partial \theta}(\sin\theta\frac{\partial}{\partial \theta})+\frac{1}{\sin^2\theta}\frac{\partial^2}{\partial \varphi^2}
\end{equation} 
is the Laplacian on the sphere.

Since $(\frac{D}{r}+\tilde{\gamma}^r\partial_r)$ is anti-hermitian we can write the eigenvalues of ${\cal D}_3$ as $\mu  + i \beta$, with $\beta$ real. The corresponding eigenfunctions, 
\begin{equation}
{\cal D}_3\,\psi_{\beta,\nu}=(\mu + i \beta) \psi_{\beta,\nu} \label{ttt}\,,
\end{equation}
have the general form
\begin{equation}
\psi_{ \beta,\nu}=f(r)\Phi_\nu(\theta,\phi)+g(r)\tilde{\gamma}^r\Phi_\nu(\theta,\phi)\,.\label{psi1}
\end{equation}
Here $\Phi_\nu(\theta,\phi)$ are the normalized eigenfunctions of the angular operator $\tilde{\gamma}^r D$, such that 
\begin{equation}
\tilde{\gamma}^r D\Phi_\nu=\nu\, \Phi_\nu\,, \hspace{1.5cm} \int d\Omega\, \Phi_\nu^\dagger \Phi_{\nu^\prime}=\delta_{\nu ,\, \nu^\prime}\,.
\end{equation}
 Since $\tilde{\gamma}^rD=(\tilde{\gamma}^rD)^{\dagger}$ the eigenvalues $\nu$ are real.  
From eqs. (\ref{h2}) it also follows that for each eigenvector $\Phi$ with eigenvalue $\nu$ of this operator, there is another eigenvector $\tilde{\gamma}^r \Phi$ with eigenvalue $2-\nu$. The spectrum of $\tilde{\gamma}^rD$ is then symmetric around the point $1$, and without loss of generality we can take $\nu \ge 1$ in equation (\ref{psi1}).

The equation (\ref{ttt}) reduces to two coupled ordinary differential equations for the radial functions $f(r)$ and $g(r)$ 
\begin{eqnarray}
\frac{f(r)}{r}\, \nu+f^{\prime}(r)&=& i \beta\, g(r)\,,\\
\frac{g(r)}{r}\,(2-\nu)+g^{\prime}(r)&=& i \beta\, f(r)\,.
\end{eqnarray}
The solutions are given in terms of Bessel functions. Keeping only the solutions for $f(r)$ and $g(r)$ which satisfy the condition (\ref{doseta})
  we arrive at the normalized eigenfunctions
 \begin{eqnarray}
&&\psi_{\beta,\nu}= i \frac{\beta}{\sqrt{|\beta|}}  \frac{J_{\nu-1/2}(r |\beta|)}{\sqrt{2 r}}   \,\Phi_\nu(\theta,\phi)+\sqrt{|\beta|}\frac{J_{\nu-3/2}(r |\beta|)}{\sqrt{2 r}}\,\tilde{\gamma}^r\Phi_\nu(\theta,\phi)\,,\label{psi}\\
&&\int_0^\infty dr\, r^2 \int d\Omega \, \psi_{ \beta^\prime,\nu^\prime }^\dagger \, \psi_{ \beta,\nu}= \delta_{\nu ,\, \nu^\prime} \delta(\beta-\beta^\prime)\,.
\end{eqnarray}  
Here we have made use of the orthogonality relation for the Bessel functions 
\begin{equation}
\sqrt{\beta \beta^{\prime}}\,\int_0^\infty dr\, r J_{s}(r \beta)  J_s(r \beta^\prime) = \delta (\beta-\beta^\prime)\,,
\end{equation}
 which is valid for $s>-1/2$.

Then, the Green function writes
\begin{equation}
G_D^{(3)}(r^\prime,\theta^\prime,\varphi^\prime,r,\theta,\varphi)=\sum_{\nu>1} \int_{-\infty}^\infty d\beta\, \psi_{\beta , \nu} (r^\prime,\theta^\prime,\varphi^\prime)\,  (\mu+i\beta)^{-1} \, \psi_{\beta , \nu}^\dagger(r, \theta, \varphi) \,.
\end{equation} 
The trace is
\begin{eqnarray}
\textrm{tr}\, G^{(3)}_D&=&\sum_{\nu>1}\int_0^\infty dr\, r \int_{0}^\infty d\beta\, \frac{\mu \beta }{ (\mu^2+\beta^2)} \left(  J_{\nu-1/2}^2(r \beta)+  J_{\nu-3/2}^2(r \beta)\right)\nonumber \\
&=&\sum_{\nu>1}\mu\int_0^\infty dr\, r  \,\left( I_{\nu-1/2}(\mu\, r)\,K_{\nu-1/2}(\mu\, r)+ I_{\nu-3/2}(\mu\, r)\,K_{\nu-3/2}(\mu\, r)\right)\,.\label{bbbh}
\end{eqnarray}
These integrals do not converge. However, here we are interested in the dependence of $\textrm{tr}\, G^{(3)}_D$ on the angle $x$ and $a$, which is finite and calculable. In other words, (\ref{bbbh}) just requires a subtraction of a global constant. In fact the integrals are divergent because the function $x \, I_\nu(x)\, K_\nu(x)$ goes to $1/2$ for $x\rightarrow \infty$, and for any $\nu$. Subtracting this global constant inside the integrals leads to 
\begin{equation}
\textrm{tr}\, G^{(3)}_D=-\mu^{-1}\sum_{\nu >1} (\nu-1)\,.
\end{equation}
This sum over the angular modes is also non convergent. We subtract a global constant by conveniently defining $\textrm{tr}\, G^{(3)}_D=0$ on $x=\pi$. At this angle there is no vertex and the logarithmic contribution vanishes.

In the appendix A we show that there are no eigenvalues $\nu$ in the interval $(1/2,3/2)$. Since the spectrum of values of $\nu$ is symmetric around $1$, we use this fact to rewrite the trace as the following sum over all the angular eigenvalues 
\begin{equation}
\textrm{tr} \, G^{(3)}_D=-\frac{1}{2 \mu}\sum_{\nu}\left|\nu-1\right|=-\frac{1}{2 \mu}\sum_{\nu}\left|\nu-1/2\right|=-\frac{1}{2 \mu} \textrm{tr}\left|\tilde{\gamma}^rD-1/2\right|\,.\label{qwqw}
\end{equation}
We make this shift since the operator on the right hand side is directly related to the Laplacian,
\begin{equation}
(\tilde{\gamma}^rD-1/2)^2=-\Delta+1/4\,.\label{uncuarto}
\end{equation}
This will allow us to connect the present problem to a similar one of a scalar field in a sphere previously studied in \cite{anguloboson}.

Note that by dimensional arguments the logarithmic divergent contribution to the partition function for an infinite plane angular sector must be proportional to  $\log (\epsilon \mu)$. This means by (\ref{pero}) and (\ref{qwqw}) that the sought logarithmic contribution to $\log Z$ is
\begin{equation}
\log Z|_{\log}=-\frac{1}{2} \textrm{tr}\left|\tilde{\gamma}^rD-1/2\right| \log (\epsilon \mu)\,.\label{questa}
\end{equation}

The trace in (\ref{questa}) can be calculated using the integral representation in terms of the resolvent \cite{representation} 
\begin{equation}
\textrm{tr}\left|\tilde{\gamma}^r D-1/2\right|=-\frac{1}{\pi}\int^{\infty}_{-\infty}dm\,m\, \textrm{tr}(i(\tilde{\gamma}^rD-1/2)+m)^{-1}\,.\label{refe}
\end{equation}
The operator 
\begin{equation}
{\cal D}_2=i(\tilde{\gamma}^r D-1/2)+ m\label{dira}
\end{equation}
 is a two dimensional Dirac operator on the sphere where the parameter $m$ plays the role of a mass. In the next Section we find the trace of the Green function $G_D^{(2)}={\cal D}_2^{-1}$, which is required by (\ref{refe}). 

\section{Green function on a sphere with a cut}
In the dimensionally reduced problem we have to find the trace of the Green function of a Dirac field on a two-dimensional sphere. This  
satisfies
\begin{equation}
{\cal D}_2 G_D^{(2)} =\sqrt{g}\, \,\delta^2(z-z^\prime)\,.
\end{equation}
The following boundary conditions for the spinors are imposed on a segment of a great circle (we choose this later on the equator)
\begin{equation}
\lim_{\varepsilon\rightarrow 0^+} \Psi(\pi/2+\varepsilon , \varphi)= e^{i 2 \pi a}  \lim_{\varepsilon\rightarrow 0^+} \Psi(\pi/2-\varepsilon , \varphi)\,,  \hspace{1.3cm} \varphi \in [\varphi_1, \varphi_2]\,.\label{boun}
\end{equation} 
We leave the end points of the cut $L_1=(\pi/2,\varphi_1)$ and $L_2=(\pi/2,\varphi_2)$ free, and at the end we will put $\varphi_2-\varphi_1=x$. 
The same boundary conditions are satisfied by the field derivatives.

The Dirac operator (\ref{dira}) on the sphere can be written\footnote{The Dirac operator in curved space is $\gamma^\mu (\partial_\mu+\Gamma_\mu)+m$. The spin connection $\Gamma_\mu$ is such that the covariant derivatives of the $\gamma^\nu$ are zero,
$D_\mu \gamma^\nu=\partial_\mu \gamma^\nu+ \Gamma^\nu_{\mu \alpha} \gamma^\alpha + [\Gamma_\mu, \gamma^\nu ]=0$.
Here we can further determine it by imposing that $\gamma^\mu \Gamma_\mu$ is diagonal. This gives the operator $\gamma^\theta \partial_\theta+\gamma^\phi \partial_\phi -i+m$, which differs from (\ref{to}) by a constant term $i/2$. As noted in the previous Section we have chosen to make this shift because in this way our fermionic operator is related to a scalar with a real mass. The ordinary Dirac operator is related to a complex mass scalar, and in this case we can not use the results of \cite{anguloboson} in a direct form (this case could also be treated with some obvious modifications of that work). 
See \cite{elabora} for further elaborations regarding the dimensional reduction of the Dirac equation in three dimensions in polar coordinates. Note also that the present problem has been solved for the hyperbolic plane where the partition function of a Dirac field with the boundary conditions analogous to (\ref{boun}) is expressed in terms of Painlev\'e functions \cite{do}.}

\begin{equation}
{\cal D}_2=\gamma^\theta \partial_\theta+\gamma^\phi \partial_\phi -\frac{i}{2}+m\,.\label{to}
\end{equation}
The gamma matrices on the sphere $\gamma^{\alpha}=i\tilde{\gamma}^{r}\tilde{\gamma^{\alpha}}$  satisfy
\begin{equation}
\{\gamma^\alpha , \gamma^\beta \}=2 g^{\alpha \beta} \,,
\end{equation}
and $g^{\theta \theta}=1$, $g^{\phi \phi}=\csc^2(\theta)$, $g^{\theta \phi}=g^{\phi \theta}=0$ are the components of the metric tensor. 
Explicitly they are
\begin{equation}
\gamma^{\theta}=\left(
\begin{array}{cc}
0 & i e^{-i\phi} \\
-i e^{i\phi} & 0
\end{array}
\right)
\,, \hspace{1cm}
\gamma^{\phi}=\left(
\begin{array}{cc}
-1 &  e^{-i\phi} \cot (\theta) \\
 e^{i\phi} \cot(\theta)& 1
\end{array}
\right)
\,. 
\end{equation}
The adjoint operator reads
\begin{equation}
{\cal D}^\dagger_2=-(\gamma^\theta \partial_\theta+\gamma^\phi \partial_\phi -\frac{i}{2})+m\,,
\end{equation}
and we have from (\ref{uncuarto})
\begin{equation}
{\cal D}_2\,{\cal D}_2^\dagger=-\Delta_\Omega +\frac{1}{4}+m^2\,. \label{dd}
\end{equation} 

\subsection{Relation between the scalar and Dirac Green functions}
Consider a complex scalar field with mass $M$ on a sphere with a cut where the boundary conditions analogous to (\ref{boun}) are imposed. The Green function $G_S^{(2)}(z,z^\prime)$ satisfies the following equations
\begin{eqnarray}
(-\Delta_{\Omega}+M^2)_z \,G_S^{(2)}(z,z^{\prime})&=&\sqrt{g}\, \delta^2(z-z^{\prime})\,,\\
\lim_{\varepsilon\rightarrow 0^+} G_S^{(2)}\left(z,(\pi/2+\varepsilon ,\phi)\right)&=&e^{i 2 \pi a}  \,\lim_{\varepsilon\rightarrow 0^+} G_S^{(2)}\left(z,(\pi/2-\varepsilon,\phi)\right)\,, \hspace{1cm} \varphi \in [\varphi_1, \varphi_2]\, .\label{cator}
\end{eqnarray} 
In order to explicitly relate this problem to the fermionic case discussed in the previously, we define an auxiliary quantity $\tilde{G}$ as
\begin{equation}
\tilde{G}(z,z^\prime)={\cal D}^\dagger_z G_S^{(2)}(z,z^\prime)\,.
\end{equation}
Eq. (\ref{dd}) implies that 
\begin{equation}
{\cal D}_z \tilde{G}(z,z^\prime) =\sqrt{g}\, \,\delta^2(z-z^\prime)\,,
\end{equation}
where the scalar and the fermion masses are related by
\begin{equation}
M^2=\frac{1}{4}+m^2\,.
\end{equation}

Thus, the difference \begin{equation}
G_D^{(2)}(z,z^\prime)-\tilde{G}(z,z^\prime)=Q(z,z^\prime)
\end{equation}
 satisfies the Dirac, 
 \begin{equation}
{\cal D} Q(z,z^\prime)=0\,,\label{didi}
\end{equation}
 and Helmholtz equation, $(-\Delta_{z}+M^2)Q(z,z^\prime)=0$, without sources.  Therefore it would be identically zero if it where bounded. $Q(z,z^\prime)$ is however unbounded at the extreme points $L_1$ and $L_2$ of the cut. Our strategy is to find a function of $z$ which also satisfy the Helmholtz equation without sources and have the same singularities as $Q(z,z^\prime)$ for $z$ going to $L_1$ and $L_2$. Then $Q(z,z^\prime)$ must be identical to this function by the uniqueness of the solution of the Helmholtz equation for bounded functions. 
 
 The singular behavior of $Q(z,z^\prime)$ (as a function of $z$) is due to two different sources. The first one is because the scalar function  $G_S^{(2)}(z,z^\prime)$ is bounded at the cut, but the leading terms go as $[(\varphi-\varphi_1)+i(\theta-\pi/2)]^{1-a }$ and $[(\varphi-\varphi_1)-i(\theta-\pi/2)]^{a }$ for $z\rightarrow L_1$, and $[(\varphi_2-\varphi)+i(\theta-\pi/2)]^{1-a }$ and $[(\varphi_2-\varphi)-i(\theta-\pi/2)]^{a }$ for $z\rightarrow L_2$. Thus, their derivatives in $\tilde{G}(z,z^\prime)$ are unbounded at these points.  
 Using the results of previous works \cite{anguloboson,boson}, these singular terms in $Q(z,z^\prime)$ are computed straightforwardly in terms of functions $S_1(z)$, $S_2(z)$, and the reflected ones, $S_1(Rz)$, $S_2(Rz)$, where $R$ is the reflexion operator which maps $L_1$ in $L_2$
\begin{equation}
R(\theta,\phi)=(\theta,\varphi_1+\varphi_2-\varphi)\,.
\end{equation}
Here $S_1(z)$ and $S_2(z)$ are uniquely defined by two requirements. They satisfy the homogeneous equation
\begin{equation}
(-\Delta_\Omega+M^2) S_1(z)=0\,,\hspace{1cm} (-\Delta_\Omega+M^2) S_2(z)=0\,,
\end{equation}
and diverge as $z\rightarrow L_1$ as
\begin{equation}
S_1(z)\sim \frac{1}{4\pi a}
[(\varphi-\varphi_1)-i(\theta-\pi/2)]^{-a } 
 \,, \hspace{1cm} S_2(z)\sim \frac{1}{4\pi (1-a)}
[(\varphi-\varphi_1)+i(\theta-\pi/2)]^{a-1 }\,.\label{wr}
\end{equation}
Then, this first source of singularities can be taken into account by the term
\begin{eqnarray}
Q_1(z,z^\prime)=&-&i 4 \pi a (1-a) \left(\gamma^\theta\vert_{L_1} S_2(z)^* S_1(z^\prime)+\gamma^\theta\vert_{L_2} S_2(Rz)^* S_1(Rz^\prime)\right)\nonumber\\
&+&4 \pi a (1-a) \left(\gamma^\phi\vert_{L_1} S_2(z)^* S_1(z^\prime)-\gamma^\phi\vert_{L_2} S_2(Rz)^* S_1(Rz^\prime)\right)
\,.
\end{eqnarray}

The second source of singular terms in $Q(z,z^\prime)$ is that $G_D^{(2)}(z,z^\prime)$ has itself singularities according to (\ref{uneta}). These must be proportional to $[(\varphi-\varphi_1)+i(\theta-\pi/2)]^{-a }$ on $L_1$ and $[(\varphi_2-\varphi)+i(\theta-\pi/2)]^{-a }$ on $L_2$, in order to respect the boundary conditions and (\ref{uneta}) (recall that $a\in (0,1/2)$). Thus, we can always write these singular terms in $Q(z,z^\prime)$ (as a function of $z$) as a combination of $S_1(z)$ and $S_1(Rz)$,   
\begin{eqnarray}
Q_2(z,z^\prime)&=&
S_1(z)^* \left(z_0(z^\prime) 1+z_1(z^\prime) \sigma_1+z_2(z^\prime)\sigma_2+z_3(z^\prime) \sigma_3 \right)\nonumber\\
&+&S_1(Rz)^* \left(y_0(z^\prime) 1+y_1(z^\prime) \sigma_1+y_2(z^\prime)\sigma_2+y_3(z^\prime) \sigma_3 \right)\,,
\end{eqnarray}
with unknown coefficient functions $z_i(z^\prime)$ and $y_i(z^\prime)$.

Therefore we have 
\begin{equation}
Q(z,z^\prime)=Q_1(z,z^\prime)+Q_2(z,z^\prime)\,.
\end{equation}   
Then, the functions $z_i(z^\prime)$ and $y_i(z^\prime)$ can be obtained in terms of the functions $S_i(z^\prime)$ inserting the above expression for $Q(z,z^\prime)$
in the Dirac equation (\ref{didi}), and using the differential equations for $S_1(z)$ and $S_2(z)$ given in \cite{anguloboson} (eqs. (81-84) of that work).
Here we need only the part of the trace of $G_D^{(2)}= \tilde{G}+ \, Q$ which is odd in the mass $m$, since the terms even in $m$ do not contribute to the integral (\ref{refe}).  After some algebra we find
\begin{equation}
  \left. \textrm{tr}\,G^{(2)}_D\,(x,m,a)\right|_{\textrm{odd}}=2 m\, \textrm{tr}\, G_S^{(2)} - \frac{16 \pi a(1-a) m\left(4\, \beta_1\,X_1  \,\cos (x/2)  - b\,B_1 \,{\sin^2 (x)} \right)}{M\,\left( 4\,{{{\beta }_1}}^2 -b^2 \sin^2 (x)  \right)} 
\,.  \label{treta}
\end{equation}
Here $\textrm{tr} \,G_S$, $b$, $\beta_1$, $X_1$ and $B_1$ are functions of $x$ which were studied in \cite{anguloboson}. They are given in terms of a system of algebraic and ordinary differential which we include in the Appendix B. 

\section{Results and conclusions}
Now we can write the formula for the coefficient of the  logarithmic contribution to the entropy due to the vertices for a Dirac field in three dimensions. The result follows combining  (\ref{vein}), (\ref{questa}) and (\ref{refe}). We have  
\begin{equation}
s_D(x)= \int_0^\infty dt \, \frac{1}{2\,\sinh^2(\pi t)}  \int_{-\infty}^{\infty}dm\,m\,\,\textrm{tr} \,G_D^{(2)}(x,m,-i t)\,.\label{aca}
\end{equation}
The relevant part of $\textrm{tr} \,G_D(x,m,-i t)$ is given by (\ref{treta}) and the formulae at the Appendix B, where we have to make the replacement $a=-i t$. The imaginary part cancel in (\ref{treta}) as it should.
  
\begin{figure} [tbp]
\centering
\leavevmode
\epsfysize=6.5cm
\bigskip
\epsfbox{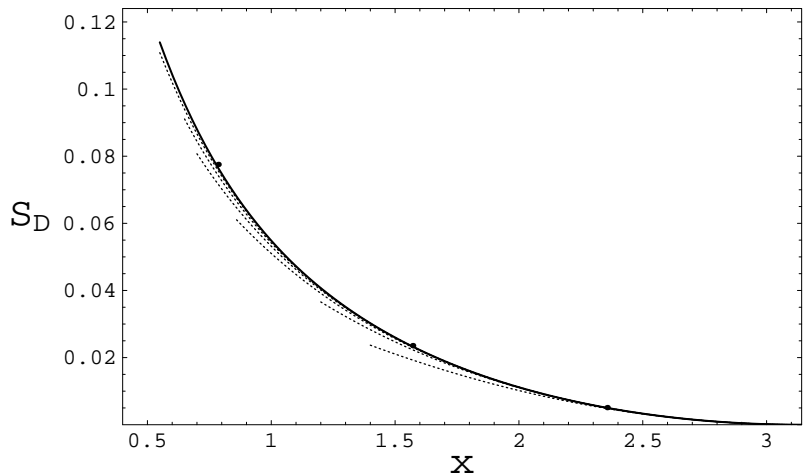}
\caption{The coefficient $s_D(x)$ of the logarithmic term in the entropy for a Dirac fermion computed by its Taylor series expansion around $x=\pi$ up to order $14$ (solid curve). The dashed lines are the Taylor expansions of lower order, from order $2$ to $12$. Also shown are three points of the curve at angles $\pi/4$, $\pi/2$, and $3/4 \pi$, obtained by direct numerical evaluation of $s_D(x)$ in a lattice (see the text for details).}
\end{figure}
 
The analogous result for a complex scalar field follows from \cite{anguloboson} with the help of the analytic continuation described in \cite{analitic}. We have 
\begin{equation}
s_S(x)=\int_0^\infty dt \, \frac{2}{\cosh^2(\pi t)} \int_{1/2}^{\infty}dM\,M\,\left(M^2-1/4\right)^{\frac{1}{2}}\, \textrm{tr} \,G_S^{(2)}(x,M,-i t+1/2)\,,
\label{final}
\end{equation}
where we have to replace $a=-i t+1/2$ in the equations (\ref{quince}-\ref{treintaydos}) of Appendix B.

An economic way to numerically integrate the equations (\ref{aca}) or (\ref{final}) is to expand the functions involved in (\ref{hprima}-\ref{treintaydos}) in Taylor series around $x=\pi$, and obtain analytically the coefficients using the differential equations. Then the above integrals over $t$ and the mass can be done for each coefficient separately. With this method we have produced the curves of figure 1, which show $s_D(x)$ up to order $(x-\pi)^{14}$. In the picture are also plotted the values of $s_D$ for $x=\pi / 4$, $\pi /2$ and $3/4 \,\pi$ obtained by numerical simulations in the lattice. They show a perfect accord (around one percent error) with the analytical results. These particular values of the angle are the ones for which the coefficient can be calculated in absolute terms with very small error on a square lattice of limited size (in the present case it was $100\times 200$ points). The numerical methods consist of evaluating the entropy for a massless Dirac field  (see \cite{fermion,chung}) for a given shape (square, triangle, etc.) and different overall size $\lambda$, and then fitting the result as $S=C_0+C_1\, \lambda+C_{-1}\, \lambda^{-1} +C_{-2} \,\lambda^{-2}- s_D \log(\lambda)$. It is also possible to evaluate  very accurately $s_D$ for specific combinations of angles using rectangular triangles.  We also obtain in this case a perfect accord with the analytical results.

\begin{table}[b]
\centering
\begin{tabular}{|c|c|c|c|c|c|c|c|} \hline
 & $c^{(\pi)}_2$ & $c^{(\pi)}_4$ & $c^{(\pi)}_6$ & $c^{(\pi)}_8$ & $c_{-1}^{(0)}$ & $s(\pi/2)$ & $s(3/4 \pi)$   \\  \hline
$S_S$ & $7.81253\, 10^{-3}$ & $5.45402 \,10^{-4}$ & $5.34656 \,10^{-5}$ & $5.40167 \,10^{-6}$ & $7.94 \, 10^{-2}$  & $0.02366$ & $0.005040$\\\hline
$S_D$ & $7.81253 \,10^{-3}$ & $5.01426\, 10^{-4}$ & $4.81299\, 10^{-5}$ & $4.85523\, 10^{-6}$ & $7.22 \, 10^{-2}$ & $0.02329$ & $0.005022$  \\  \hline
$S_H$ & $7.81253 \,10^{-3}$ & $4.94734\, 10^{-4}$ & $4.63675 \,10^{-5}$ & $4.64246\, 10^{-6}$ & $7.04\, 10^{-2}$  & $0.02321$ & $0.005019$ \\ \hline
\end{tabular} 
\caption{The first four non zero Taylor coefficients of $s_D(x)$, $s_S(x)$ (a complex scalar) and $s_H(x)$ for $x$ around $\pi$, $s(x)\sim \sum c^{(\pi)}_j\, (x-\pi)^j$,  and the coefficient of the term $1/x$ of these functions for $x\rightarrow 0$, $s(x)\sim  c_{-1}^{(0)}/x$. The value of the functions for $x=\pi/2$ and $x=3/4\, \pi$ are also shown.  $s_H(x)$ is normalized such that the quadratic coefficient coincide with the ones of $s_D(x)$ and $s_S(x)$. }
\end{table}

In \cite{taka} Ryu and Takayanagi propose a purely geometric method to compute the entanglement entropy for certain conformal field theory in the AdS-CFT context. In this way Hirata and Takayanagi find a logarithmic contribution to the entropy due to the vertices in three dimensions which presumably corresponds to an interacting CFT \cite{angholo}. The coefficient of the logarithmic term $s_H(x)$ in this holographic theory is given in parametric form in terms of a variable $g_0\in (0,\infty)$ as
\begin{eqnarray}
x(g_0)&=& 2 g_0 \sqrt{1+g_0^2}\int^\infty_0 \frac{dz}{(z^2+g_0^2) \sqrt{(z^2+g_0^2+1)(z^2+2 g_0^2+1)}}\,,\\
s_H(g_0)&=& N \, \int^\infty_0 dz\,\left(1-\sqrt{\frac{z^2+g_0^2+1}{z^2+2 g_0^2+1}}\right)\,.\label{jaka}
\end{eqnarray}

\begin{figure} [tbp]
\centering
\leavevmode
\epsfysize=6cm
\bigskip
\epsfbox{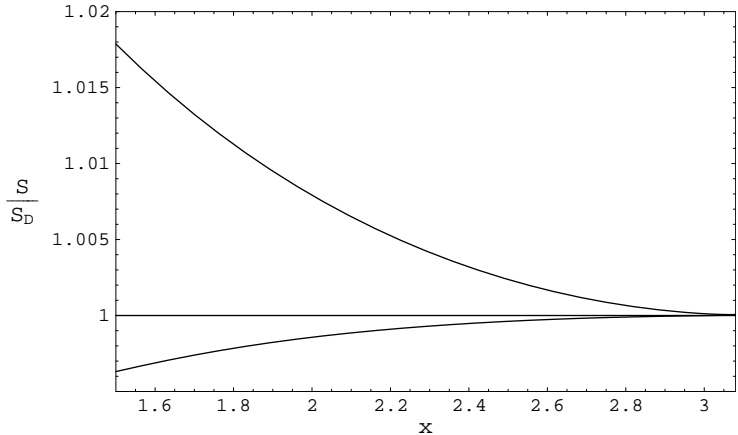}
\caption{Comparison between the coefficients of the logarithmic term for different theories. The picture shows, from top to bottom, the functions $s_S$, $s_D$ and $s_H$ normalized to $s_D$. The differences increase toward the origin where these ratios get $s_S/s_D=1.099$ and $s_H/s_D=.975$ (not shown).}
\end{figure}

In order to compare the functions $s_D(x)$, $s_S(x)$ and $s_H(x)$ we have to choose a normalization $N$ for $s_H(x)$ in (\ref{jaka}). Remarkably, the quadratic coefficients of $s_D(x)$ and $s_S(x)$ turn out to be equal. We have then chosen to normalize $s_H(x)$ such that its quadratic coefficient coincides with the ones of $s_D(x)$ and $s_S(x)$. In Table 1 we show the first four non zero Taylor coefficients of $s_D(x)$, $s_S(x)$ and $s_H(x)$ for $x$ around $\pi$. 

As explained in \cite{anguloboson}, all of these functions diverge as $1/x$ for $x$ going to zero. In the Table 1 we also show the coefficient of $1/x$ in the expansion around $x=0$. For $s_D(x)$ and $s_S(x)$ this small angle behavior is given by \cite{anguloboson} 
\begin{equation}
s(x)\sim \frac{\int_0^\infty dt\, C(t)/\pi}{x}\,,\label{utut}
\end{equation} 
where $C(t)$ is the entropic c-function corresponding to a Dirac field and complex scalar in two dimensions respectively \cite{irrev}. The integral in (\ref{utut}) is done numerically using the analytic expressions for $C(t)$ given in \cite{analitic}.

In figure 2 we have plotted $s_S(x)/s_D(x)$ and $s_H(x)/s_D(x)$ for a range $x$ where our integration of these functions is precise enough.  Clearly, they are linearly independent. This points to the effect of interactions in the CFT behind $s_H(x)$.

As a cross check of these results one can verify in each case that $s(x)$ satisfies the constraints coming from the strong subadditive inequality for the entropy. These are given by
\begin{eqnarray}
s(x)&\ge&  0\,, \label{tal}\\
s^\prime(x)&\le& 0 \,,\\
 s^\prime(x)+ \sin(x)\, s^{\prime\prime}(x)&\ge& 0\,,\label{cual}
\end{eqnarray} 
where $x\in (0,\pi]$. They are obtained by considering the strong subadditive inequality for different spatial (but relatively boosted to each other \cite{noper}) plane angular sectors with a common vertex in Minkowski space. This problem can be mapped to a two dimensional de-Sitter space by taking the intersection of the plane angular sectors with an hyperboloid $x^\mu x_\mu=-1$. Then, the inequalities (\ref{tal}-\ref{cual}) for the two dimensional de-Sitter case follow in a similar way as in the flat space case discussed in \cite{irrev} (see also \cite{angholo} where a weaker form of (\ref{cual}) was derived).

The coefficient of $\log (\epsilon)$ does not depend on the mass. Therefore we conclude that there are conformal field theories in three dimensions with non proportional entropy functions. This is in contrast to the $1+1$ case, where all the CFT would have entanglement entropy functions which are proportional to a unique function of the set $V$ \cite{see,para} (see however the recent works in \cite{howe}, and \cite{irrev}).

However, note that all the functions are remarkably similar to each other. There is a maximal relative difference between the scalar and Dirac case of $9\%$ (the maximum relative distance seems to happen for $x\rightarrow 0$). The Hirata-Takayanagi formula approaches better the fermion than the boson case, with a maximum relative difference with respect to $S_D$ (at the origin) which is remarkably small: only $2.5\%$.  

\section*{Acknowledgements}
H.C. and M.H. thank CONICET, ANPCyT and UNCuyo for financial support.

\section*{Appendix A: Gap in the spectrum of the Dirac operator on the sphere with boundary conditions specified on a segment of a great circle}
In Section 2 we used that the operator $\tilde{\gamma}^r D$ with boundary conditions on a segment of a great circle given by (\ref{boun}) has no eigenvalues $\nu$ in the interval $(1/2,3/2)$. In this Appendix we provide the proof of this fact. Since the spectrum is symmetric around $\nu=1$ we have to show there are no eigenvalues in $(1/2,1]$. First we note that due to (\ref{h2}) the eigenvectors $\Phi_\nu$ of $\tilde{\gamma}^r D$ corresponding to eigenvalue $\nu$ satisfy 
\begin{equation}
\Delta \, \Phi_\nu= -\nu (\nu-1) \, \Phi_\nu\,.\label{lapla}
\end{equation}
For $a=0$ the functions $\Phi_\nu$ are regular and (\ref{lapla}) implies due to the positivity of $-\Delta$ for regular functions that $\nu\notin (0,1)$. The case $\nu=1$ is also not possible since in this case the spinor only contains eigenvectors of $\Delta$ with $0$ eigenvalues and thus it is a constant spinor, giving $\nu=0$ instead. 

Now, as we increase $a$ from $0$ to $1/2$ the eigenvalues start moving. In order that at some point an eigenvalue enters to $ (1/2,1]$ it must cross the point $\nu=1/2$. At $\nu=1/2$ we have 
\begin{equation}
\Delta  \, \Phi_{1/2}= \frac{1}{4} \, \Phi_{1/2}\,,\label{eye}
\end{equation}   
and the spinor diverges at the extreme points of the cut as $d^{-1/2}$ (see eq.  (\ref{uneta})). This means, according to \cite{anguloboson} (see also Section 3 in this paper), that it is possible to write
\begin{equation}
\Phi_{1/2}(\theta,\varphi)=\left( 
\begin{array}{c}
	m_{1,1}\,S_1(\theta,\varphi) +m_{1,2}\, S_1(R(\theta,\varphi)) \\m_{2,1}\,	S_1(\theta,\varphi) +m_{2,2}\, S_1(R(\theta,\varphi))\label{epat}
\end{array}
\right)\,,
\end{equation}
where the $m_{i,j}$ are unknown constant coefficients and the $S_1(\theta,\varphi)$ and $S_1(R(\theta,\varphi))$ are the functions introduced in \cite{anguloboson} and described also in Section 3. These must satisfy the eigenvalue equation (\ref{eye}) and thus correspond to mass $M=1/2$. If $\nu=1/2$ is eigenvalue of $\tilde{\gamma}^r D$ the equation (\ref{epat}) has to hold for some coefficients, since adequately choosing them we can equate the divergent contributions on both sides of the equation. Then the difference between the left and right hand sides satisfies the Helmholtz equation $(\Delta-M^2)=0$, and is singularity free, what means it is identically zero.

Inserting (\ref{epat}) in the eigenvalue equation for $\tilde{\gamma}^r D$, and using the relevant differential equations for $S_1(\theta,\varphi)$ and $S_1(R(\theta,\varphi))$ (see \cite{anguloboson}) we arrive at
\begin{equation}
 -u\, \beta_1 + (2 a-1) b\, \sin(x/2)=0\,.
\end{equation}
Here $u$, $\beta_1$ and $b$ are functions of $a$ and $x$ which satisfy the differential equations given in appendix B (for $M=1/2$). With these one readily shows that there are no solutions for $a\in (0,1/2)$.   

\section*{Appendix B: The Green function of a scalar field in a sphere with boundary conditions specified on a segment of a great circle}
In order to make this paper self contained we include the results of \cite{anguloboson} on the Green function of a complex scalar field on a sphere with the boundary conditions (\ref{cator}). The trace is
\begin{equation}
\textrm{tr}\,G_S(x,M,a)=8\pi (1-a)a\int_{x}^{\pi}H_a(y)dy\,.\label{quince}
\end{equation}
The function $H_a(x)$ is the solution of the following set of ordinary non linear differential equations (we omit the subscript $a$ and the dependence on $x$ and the mass $M$ of the variables for notational convenience)
\begin{eqnarray} 
H' &=& - \frac{M}{2}\,\left( b\,B_2 +c\, B_1 + 2\,u\,B_{12} \right)\label{hprima}\,,\\
X_1' &=& - M\,\left( b\,B_{12}+ u\,B_1 \right)\label{x1prima} \,,\\
X_2' &=& - M\,\left( c\,B_{12} + u\,B_2 \right) \label{x2prima}\,,\\
c' &=&- 2\,M\,\beta_2 \,u\, \csc(x)\,\sin(x/2) - c\,(1 - a)\,\csc(x)\,(1 + \cos(x))\label{cprima}\,, \\
b' &=& -2\,M\,\beta_1\,u\,\csc(x)\,\sin(x/2) - b\,a\,\csc(x)\,(1 + \cos(x)) \label{bprima} \,,\\
u'&=&-\frac{M}{2}\,\sec(x/2)\,(b\,\beta_2+ c\,\beta_1) + \frac{1}{2}\,u\,\tan \left(x/2 \right)\label{uprima} \,,
\end{eqnarray}
where $B_1$, $B_2$, $B_{12}$, $\beta_1$, $\beta_2$ are functions of $x$ given in terms of $H$, $X_1$, $X_2$, $c$, $b$, and $u$ by the following set of algebraic equations
\begin{eqnarray}
\frac{\cos(x/2)}{8 
\pi a(1 - a)}&=&\sin(x/2)\,H - M \left(\beta_1\, X_2 + \beta_2\, X_1\right) + 2\,M\,\cos(x/2)\, u\, B_{12}\,,\label{29} \\
\frac{\sin(x/2)}{8 \pi a(1 - a)}&=&-\cos(x/2)\,H - M\, \tan(x/2)\left(\beta_1 X_2 + \beta_2 X_1 \right) + 
  M\,\sin(x/2)(b B_2 + c B_1) \,,\label{30}\\
0&=& -M \sin(x/2)(c X_1
 -b X_2) + M \tan(x/2)(\beta_2 B_1- \beta_1 B_2)+(1-2a)\cos(x/2) B_{12}\,,\label{31}\\
0&=&-4a(a - 1) - M^2(4 - 8\beta_1\beta_2 + b c + 3u^2) \nonumber\\&&\hspace{2cm}- 
      4\cos(x)\left(a(a - 1) + M^2(u^2 + 1)\right) 
      + M^2\cos(2x)(b\,c - u^2) \label{4a}\,,\\
0&=&(2a - 1)u\cos(x/2)+M \tan(x/2)(\beta_1 c - b \beta_2)\,.\label{minuscula}
\end{eqnarray}
The boundary conditions at $x\rightarrow\pi$ are
\begin{eqnarray}
H(\pi)&=&0\,,\label{hache}\\
X_1(\pi)&=&\frac{ \Gamma(-a) \left( \cosh \left( \frac{\pi \mu_1}{2} \right) \textrm{Im} \left[ \psi \left( \frac{1}{2} + a + \frac{i\mu_1}{2} \right) \right] - \frac{\pi}{2} \sinh \left( \frac{\pi\mu_1}{2} \right)   \right)}{2^{2a}\mu_1  \left( \cos \left( 2 a \pi\right)+\cosh (\pi \mu_1)\right) \Gamma (1+a) \left| \Gamma \left( \frac{1}{2}-a+\frac{i\mu_1}{2}\right) \right|^2 }\,,\label{xx1}\\
X_2(\pi)&=&X_1(\pi)\left|_{\,a\rightarrow (1-a)} \right. \,,\label{xx2} \\
u(\pi)&=&0\,, \label{upi}\\
b	(\pi)&=&\frac{ 2^{1-2a} a (1-a) \left| \Gamma\left( \frac{1}{2} +a+\frac{i\mu_1}{2}\right)\right|^2}{M\Gamma^2 (1+a)}\,,\label{treintaytres}\\
c(\pi)&=&b(\pi)\left|_{\,a\rightarrow (1-a)}\right.  \,,\label{treintaydos}
\end{eqnarray}
where $\mu_1=\sqrt{4\,M^2-1}$ and $\psi$ is the digamma function. The meaning of the extra variables $B_1$, $B_2$, $B_{12}$, $X_1$, $X_2$, $u$, $b$, $c$, $\beta_1$ and $\beta_2$ is the same as in \cite{boson}. The trace in (\ref{quince}) is regularized such that it vanishes when $x=\pi$, where there is no vertex point and no logarithmic term is present in the entropies.

\end{document}